\documentclass[journal]{IEEEtran}
\ifCLASSINFOpdf
\else
\fi
\usepackage{graphicx}
\usepackage{amssymb}
\usepackage{bbm}
\usepackage{mathrsfs}
\usepackage{amssymb}
\usepackage{amsmath}
\usepackage{amsthm}
\usepackage{latexsym}
\usepackage{indentfirst}
\usepackage{enumitem}
\usepackage{anysize}
\usepackage{array}
 \usepackage[colorlinks=true]{hyperref}
\hypersetup{urlcolor=blue, citecolor=red}
\begin{document}
\newcommand{\bbF}{\mathbb{F}}
\newcommand{\bbZ}{\mathbb{Z}}
\newcommand{\F}{\mathbb{F}}
\newcommand{\Z}{\mathbb{Z}}
\newcommand{\Q}{\overline{Q}}
\newcommand{\N}{\overline{N}}
\newtheorem{theorem}{Theorem}[section]
\newtheorem{corollary}{Corollary}
\newtheorem*{main}{Main Theorem}
\newtheorem{lemma}[theorem]{Lemma}
\newtheorem{proposition}[theorem]{Proposition}
\newtheorem{conjecture}{Conjecture}
\newtheorem*{problem}{Problem}
\theoremstyle{definition}
\newtheorem{definition}[theorem]{Definition}
\newtheorem{remark}{Remark}
\newtheorem*{notation}{Notation}
\newcommand{\ep}{\varepsilon}
\newcommand{\eps}[1]{{#1}_{\varepsilon}}%
\title{Asymptotically Good Additive Cyclic Codes Exist}
%
%
%

\author{{Minjia Shi\thanks{Minjia Shi (corresponding author) is with Key Laboratory of Intelligent Computing \& Signal Processing
Ministry of Education, Anhui University
No. 3 Feixi Road, Hefei
Anhui Province 230039, P. R. China, and School of Mathematical Sciences of Anhui University, Anhui, 230601, P. R. China (E-mail: smjwcl.good@163.com)}, Rongsheng  Wu\thanks{Rongsheng Wu is with School of Mathematical Sciences of Anhui University, Anhui, 230601, P. R. China (E-mail: wrs2510@126.com)}, Patrick Sol\'e}\thanks{Patrick Sol\'e is with CNRS/LAGA, University of Paris 8, 2 rue de la libert¨¦, 93 526 Saint-Denis, France}}
\maketitle

\begin{abstract}
Quasi-twisted codes of fixed index $>1$ have been shown recently to be asymptotically good (A. Alahmadi, C. G\"uneri, H. Shoaib, P. Sol\'e, 2017).
We use this result to construct asymptotically good  additive constacyclic codes on any extension of fixed degree of the base field. A similar result is proved for additive cyclic
codes, under Artin Conjecture on primitive roots.
\end{abstract}


%
\IEEEpeerreviewmaketitle

\section{Introduction}
%
%
%
%
\IEEEPARstart{C}YCLIC codes are one of the oldest and most important class of codes. Most classical codes are either cyclic like the Hamming codes,
the Golay codes, and the Reed-Solomon codes, or extended cyclic like the Reed-Mueller codes. They enjoy connections with shift-register sequences,
computable lower bounds on the minimum distance, and, for some subclasses, algebraic decoding techniques like BCH decoding for instance \cite{B2,HP}. However,
their asymptotic performance is still unsettled to this day. BCH codes, the most important class of cyclic codes, are asymptotically bad \cite{LP}. Cyclic codes with an affine-invariant extension are asymptotically bad \cite{K}.
 Cyclic codes with lengths having prime factors
belonging to a fixed finite set of prime numbers are asymptotically bad  \cite{Ca}.
 On the positive side, dihedral codes, a class of linear codes that looks even more constrained than cyclic codes,  have been shown to be good \cite{AOS,BM}.
Determining if cyclic codes are asymptotically good is a half-century old open problem  \cite{AM,C,EL,MW}. In contrast, quasi-cyclic codes of index $>1$ (the notion of index is to be defined below) have been known to be
asymptotically good since the 1960's \cite{CPW}.
This dichotomy can be explained intuitively by the rarity of cyclic codes of given length, which precludes the use of random coding arguments.
Thus, binary quasi-cyclic codes of index $2$ were shown to be asymptotically good in \cite{CPW}. This was extended to any finite field, and refined to self-dual codes in \cite{AOS}.
The result was further extended to any index  in \cite{AGSS}.

In the present paper, we bridge the gap between cyclic codes and quasi-cyclic codes codes by establishing a map between quasi-cyclic codes of index $\ell$ over a finite field, and additive cyclic codes over an extension of degree $\ell$. The term additive over a field $K$ means that the code is only assumed to be linear over a strict subfield $L$ of $K.$ For instance quaternary additive codes ($K=\F_4$) are only supposed to be $\F_2$-linear ($L=\F_2$).
The example of the dodecacode, a self-dual code of length $12$ and minimum distance $6$ shows that this allows for stronger codes. The best linear self-dual quaternary code has only minimum distance $4$ \cite{C+}.

 To gain in generality, we give a map between
additive constacyclic codes and quasi-twisted codes. Constacyclic codes are a natural generalization of cyclic codes when the alphabet is nonbinary. They are defined as linear codes invariant under the constashift (to be defined  below), a one-parameter generalization of the cyclic shift. They have been used to construct codes for the
Lee metric \cite{B}, and, under the name pseudo-cyclic, to build MDS codes over large alphabets \cite{PD}. They are the case of index unity of quasi-twisted codes. Thus constacyclic codes and quasitwisted codes are in the same relation as cyclic codes and quasi-cyclic codes. The class of quasi-twisted codes, which contains both constacyclic codes (when the index is one) and quasi-cyclic codes (when the constashift is the usual shift) as a subclass, forms an important class of linear codes \cite{CPW,JY,LS}.
In particular, we show that additive constacyclic codes over field extensions, a superclass of constacyclic codes, are asymptotically good, by deriving a modified Gilbert-Varshamov bound for them.
Additive cyclic codes are motivated by the construction of quantum error-correcting codes \cite{B2,BE}, and enjoy a structure theory similar to that of cyclic codes \cite{B1}.
This latter property has been extended recently to additive constacyclic codes \cite{CCC}.
We build on a natural map between quasi-twisted codes of index $\ell$ over $\F_q,$ and additive constacyclic codes over $\F_{q^\ell}.$ This map was exploited in \cite{STGSS}
when $\lambda=1$ in the reverse direction to construct quasi-cyclic codes from cyclic codes. Note that the connection with additive cyclic codes has already been observed in \cite{GOO}.
Combining this last observation with the existence result of \cite{AGSS}, we can show that additive constacyclic codes over extension field are asymptotically good.

The material is organized as follows. Section II collects standard notation and notions on codes and asymptotics. Section III studies the said map and uses it to construct additive constacyclic codes in Theorems 3.4 and 3.6.
 Section IV concludes this paper and sketches some open problems.

\section{Definitions and Notation}
Let $q$ be a prime power, and $\F_q$ denote the finite field of order $q.$ Let $\F_q^*$ denote $\F_q\backslash \{0\}$.
By a {\bf code} of length $n$ over $\F_q,$ we shall mean a proper subset of $\F_q^n.$ This code is {\bf linear} if it is a  $\F_q$-vector subspace of $\F_q^n.$
 The {\bf dimension} of a code $C$, denoted by $k$, is equal to its dimension as a vector space.
Its (minimum) {\bf distance}, denoted by $d$ or $d(C),$ is defined as the minimum Hamming weight of its nonzero elements. The {\bf Hamming weight}  of $x\in \F_q^n,$ denoted by $w(x),$ is the number of indices $i$ where $x_i \neq 0.$
The three parameters of a code are written compactly as $[n,k,d].$
We extend this notation to a possibly nonlinear code $C \subseteq \F_q^n,$ by letting then $k =\log_q(\vert C\vert),$ and letting $d$ being the minimum pairwise distance between two nonzero codewords.
If $C(n)$ is a family of codes of parameters $[n, k_n, d_n]$, the {\bf rate} $r$ and {\bf relative distance} $\delta$ are defined as $$r=\limsup\limits_{n \rightarrow \infty}\frac{k_n}{n},$$ and
\begin{equation*}\delta=\liminf\limits_{n \rightarrow \infty}\frac{d_n}{n}.\end{equation*}
A family of codes is said to be {\bf asymptotically good} iff $r\delta >0.$

Recall that the $q$-ary {\bf entropy function} $H_q(\cdot)$ is defined for $0<y< \frac{q-1}{q}$ by $$ H_q(y)=y\log_q(q-1)-y\log_q(y)-(1-y)\log_q(1-y).$$

Let $C$ be a linear code over $\F_q$ of length $n$ and $\lambda \in \F_q^*$, we define the {\bf constashift } $T_{\lambda}$ by its action on the codewords $c=(c_0,c_1,\ldots,c_{n-1})$ as:
$$T_\lambda(c)=(\lambda c_{n-1},c_0,\ldots,c_{n-2}).$$ Thus, $T_1$ is just the usual cyclic shift; $T_{-1}$ is sometimes called the {\bf negashift}.
A linear code $C$ is called {\bf $\lambda$-constacyclic} if $C$ is invariant under $T_\lambda$, i.e., $T_\lambda(C)= C$. In particular, the code $C$ is called \textbf{cyclic} if $\lambda=1$, and \textbf{negacyclic} if $\lambda=-1.$
This linear code $C$ is called a $(\lambda,\ell)$-\textbf{quasi-twisted} (QT) code if $C$ is invariant under the action $T_\lambda^\ell$, and the integer $\ell$ is called the \textbf{index} of $C$. It is easy to check that a $(\lambda,\ell)$-QT code of length $n$ is also a $(\lambda, \gcd(n,\ell))$-QT code, so we always assume that $\ell\mid n$. For simplicity we assume that $n= \ell m$ for some integer $m$, sometimes called the {\bf co-index}. The special case $\lambda= 1$ gives the more familiar class of $\ell$-\textbf{quasi-cyclic} (QC) code.

 By an {\bf additive $\lambda$-constacyclic code} over $\mathbb{F}_{q^\ell},$ we shall mean an $\F_q$-linear code over the alphabet $\mathbb{F}_{q^\ell}$
 that is invariant under the shift $T_\lambda.$ In particular, when $\lambda=1$, i.e., additive cyclic code,
 see \cite{B1,B2,BE} for background material on this important family of codes.

 For simplicity, we may write QT, QC, or constacyclic instead of $(\lambda,\ell)$-QT, $\lambda$-QC , or $\lambda$-constacyclic.
 We illustrate the link between various families of codes by the following diagram where horizontal arrow means $\lambda=1,$ and vertical down arrow means $\ell=1.$
 \[\begin{array}{ccc} QT &
{\longrightarrow} &
QC\\
\big\downarrow & &
\big\downarrow\vcenter{%
\rlap{}}\\
\lambda$-$cyclic & \longrightarrow & cyclic
\end{array}\]
\section{From $(\lambda,\ell)$-QT codes to additive $\lambda$-constacyclic codes}
Let $m$ be a positive integer satisfying $\gcd(m,q)=1$. Define $\mathbb{F}_{q}[x]$ as the ring of polynomials in the indeterminate $x$ over $\mathbb{F}_{q}$.
Consider the ring $R_{\lambda}(m,q)=\mathbb{F}_{q}[x]/(x^m-\lambda)$ with $\lambda\in \F_q^*.$ Map a codeword $(c_0,c_1,\ldots, c_{m-1})\in C$ to the polynomial $c_0+c_1x+\cdots +c_{m-1}x^{m-1} \in \F_q[x].$
This called the polynomial representation. Using this representation, it is easy to
show that a $\lambda$-constacyclic code of length $m$ is an ideal in the ring $R_{\lambda}(m,q).$
Similarly QT codes of index $\ell$ and co-index $m$ are $R_{\lambda}(m,q)$-modules \cite{JY,LS}.

In the language of polynomials, a codeword of a $(\lambda,\ell)$-QT code can be written as $c(x)=(c_{0}(x),\ldots,$ $c_{\ell-1}(x))\in R_{\lambda}(m,q)^\ell.$ Given a basis $B=\{ e_{0},\ e_{1},\ \ldots, \ e_{\ell-1} \}$ of $\mathbb{F}_{q^\ell}$ over $\mathbb{F}_{q}$, now we can define the following map:
\begin{eqnarray*}
\phi_B: R_{\lambda}(m,q)^\ell \ &\rightarrow&  R_{\lambda}(m,q^\ell),\\
(c_{0}(x),c_{1}(x),\ldots,c_{\ell-1}(x)) & \longmapsto & \sum\limits_{i=0}^{\ell-1}c_i(x)e_i.
\end{eqnarray*}
{\bf Example:} Take $q=3$ and $\ell=2.$ A quasi-twisted code of length $4=2\times 2$ and index $2,$ with $\lambda=-1,$ is the ternary Hamming code of parameters $[4,2,3]$ with generator matrix
 $$\left(\begin{array}{cccc}1 & 0 & 1 & 1\\
 0 & 1 & 1 & -1
  \end{array}\right), $$
 or $(1,1+x)$ in polynomial form \cite[Example 1.3.3., p.6]{HP}.
Its image under $\phi_B$ with $B=\{1,\iota\},$ where $\iota^2=-1$ is an additive negacyclic code over $\F_9$ generated by the $\F_3$-span of the negashifts of $(1+\iota,\iota).$

For the case $\lambda=1$, the inverse map was used in \cite{STGSS} to define a special class of quasi-cyclic codes from cyclic codes over $\mathbb{F}_{q^\ell}.$ In the present paper, we will use this map to construct additive $\lambda$-constacyclic codes over $\mathbb{F}_{q^\ell},$ from $(\lambda,\ell)$-QT codes of index $\ell$ and co-index $m$, over $\mathbb{F}_{q}.$

The following result is trivial but essential, and was observed in \cite{GOO} when $\lambda=1$.\\
{\bf Theorem 3.1}\label{mainfirst}
If $C$ is a $(\lambda, \ell)$-QT code of length $n=\ell m$ over $\mathbb{F}_q,$  then  $\phi_B(C)$ is an additive $\lambda$-constacyclic code of length $m$ over $\mathbb{F}_{q^\ell}.$
\begin{proof}
The $\F_q$-linearity of the code $\phi_B(C)$ follows by the $\F_q$-linearity of the map $\phi_B$, which is immediate by the definition above.
It is easy to check that the code $\phi_B(C)$ is invariant under the action of $T_\lambda,$ with $\lambda\in \F_q^*$. In the language of polynomials, the action of $T_\lambda$ amounts to multiplication by $x.$ Thus
$$\phi_B(x(c_{0}(x),c_{1}(x),\ldots,c_{\ell-1}(x)))=x\sum\limits_{i=0}^{\ell-1}c_i(x)e_i.$$
This completes the proof.
\end{proof}

{\bf Remarks:} 1) The image $\phi_B(C)$ of a linear code $C$ is not $\mathbb{F}_{q^\ell}$-linear in general. For instance, it is not the case if the dimension $k$ of $C$ over $\F_q,$ is not a multiple of $\ell$ since
$\vert \phi_B(C)\vert=\vert C\vert=q^k=(q^\ell)^{\frac{k}{\ell}}.$ Some necessary conditions can be found in \cite{GOS}.\\
2) The reverse map $\phi_B^{-1}$ is very well-known in Coding Theory. Thus the Golay code is the $\phi_B^{-1}$ image of a Reed-Solomon code over $\F_8$ with $B$ a polynomial basis of $\F_8$
 over $\F_2$ \cite{W}. See \cite{STGSS} for applications and references.

The rest of the section is as follows. Firstly, Lemma 3.2 controls the minimum distance of $\phi_B(C)$ as a function of that of $C.$
Secondly, Lemma 3.3 recalls the existence of asymptotically good QT codes from \cite{AGSS}. From the application of these two lemmas, and Theorem 3.1 the main result
(Theorem 3.4) will follow. Since $x^m-1$ is reducible over any field for $m>1,$ we need an alternative to Lemma 3.3 to derive  the analogue of Theorem 3.4
(that is Theorem 3.6) when
$\lambda=1.$ This requires the use of Artin conjecture. This argument requires $q$ to be a non-square, a restriction that can be removed easily (Corollary 3.7).\\

{\bf Lemma 3.2}\label{rabi}  Let $C$ be a $(\lambda,\ell)$-QT code of length $n=\ell m$ over $\F_q$, with distance $d(C)$, then we have the bound on the distance of $d(\phi_B(C))$ given by
$$ d(\phi_B(C))\ge \bigg\lceil\frac{d(C)}{\ell}\bigg\rceil.$$

\begin{proof}
Let $c=(c_0,c_1\dots,c_{\ell-1})\in C,$ with $c\neq 0,$ and with $c_i\in \F_q^m$ for all $i$'s. Put $z=\phi_B(c).$
Then $z=\sum_{i=0}^{\ell-1} c_i e_i.$ Consider $z_j$ an arbitrary component of $z.$ Thus, by linearity, $z_j=\sum_{i=0}^{\ell-1} c_{ij} e_i,$ with $c_{ij}$ component of index $j$ of $c_i.$
Since $B$ is a basis $z_j=0$ entails $c_{ij}=0$ for all $i$'s. This, in turn, proves that $\ell w(z_j)\ge \sum_{i=0}^{\ell-1}w(c_{ij}).$
But $$w(c)=\sum_{i=0}^{\ell-1}\sum_{j=0}^{m-1}w(c_{ij}),$$ and $w(z)={\sum\limits_{j=0}^{m-1}w(z_j)}.$ The result follows by summing $m$ inequalities.
\end{proof}

Now, we assume that $x^m-\lambda$ is irreducible in $\F_q[x]$. By \cite{RHN}, we know this is true for the following conditions:
\begin{enumerate}
  \item Each prime factor of $m$ divides the order $a$ of $\lambda$ in $\F_q^*$, but not $(q-1)/a$;
  \item $q\equiv 1(\rm{mod}~ 4)$ if $m\equiv 0(\rm{mod}~ 4)$.
\end{enumerate}

We will require a lemma on $(\lambda,\ell)$-QT codes. The proof is omitted.\\
{\bf Lemma 3.3} (\cite[Theorem 5.4]{AGSS})
\label{QT}
Let $q$ be a prime power, and keep the notations above such that the polynomial $x^m-\lambda$ is irreducible in $\F_q[x]$. Then for any fixed integer $\ell\ge 2,$ there are infinite families of $(\lambda,\ell)$-QT codes of length $n=\ell m,$ index $\ell$, rate $1/\ell$ and of relative distance $\delta$,
satisfying  $H_q(\delta)\geq \frac{\ell-1}{\ell}$.

We are now in a position to state and prove the main result of this section.\\
{\bf Theorem 3.4} \label{main}Let $q$ be a prime power.  There are infinite families of additive constacyclic codes of length $m\rightarrow \infty$ over $\F_{q^\ell}$ of rate $1/\ell$ and relative distance
$$\delta \ge \frac{1}{\ell}H_q^{-1}(1-1/\ell).$$
\begin{proof}
First, we keep the conditions described above such that $x^m-\lambda$ is irreducible in $\F_q[x]$, then invoking Lemma 3.3, we can construct an infinite family of $(\lambda,\ell)$-QT codes of index $\ell$ and relative distance $\delta'$ such that
$\delta' \ge H_q^{-1}(1-1/\ell).$
Next, we observe, by Lemma 3.2, that there is then an infinite family of additive constacyclic codes (by Theorem 3.1) of relative distance $\delta$ that satisfies
$\delta' \le \ell \delta.$
The result follows from the above inequalities.
\end{proof}

Note that the inverse function of $H_q(\cdot)$ is denoted by $H_q^{-1}(\cdot)$. For the special case $\lambda=1$, we give the following lemma on quasi-cyclic codes depending the Artin conjecture.\\
{\bf Lemma 3.5} (\cite[Theorem 5.6]{AGSS})
\label{QC}
Let $q$ be a prime power that is not a square, and $m$ be a prime. If $x^m-1=(x-1)u(x),$ with $u(x)$ irreducible over $\F_q[x],$ then for any fixed integer $\ell\ge 2,$ there are infinite families of QC
codes of length $\ell m,$ index $\ell$, rate $1/\ell$ and of relative distance $\delta$, satisfying  $H_q(\delta)\geq \frac{\ell-1}{\ell}$.\\

Note that, by the theory of cyclotomic cosets, there are infinitely many $n$'s satisfying the hypothesis of Lemma 3.5 iff there are infinitely many $n$'s such that $q$ is primitive modulo $n.$ This latter condition is known as Artin conjecture and
has been proved by Hooley under GRH \cite{H} when $q$ is not a perfect square (if $q$ is a square, it is never primitive for $p>2,$ since $q^{(p-1)/2}\equiv 1 \pmod{p},$ for all odd primes $p$).

We can state an analogue of Theorem 3.4 for $\lambda=1.$ The proof is similar and omitted.\\
{\bf Theorem 3.6} \label{main2}Let $q$ be a prime power that is not a square.  There are infinite families of additive cyclic codes of length $m\rightarrow \infty$ over $\F_{q^\ell}$ of rate $1/\ell$ and relative distance
$\delta \ge \frac{1}{\ell}H_q^{-1}(1-1/\ell).$

The restriction on $q$ can be removed easily.\\
{\bf Corollary 3.7} If $q$ is a prime power, then there are good families of  additive cyclic codes of length $m\rightarrow \infty$ over $\F_{q^\ell}$ of rate $ \ge 1/\ell$ and relative distance
$\delta >0.$
\begin{proof}
 Suppose $q$ is a square of the form $p^{2^a (2b+1)},$ with $a\ge 1,$ and let $s= p^{2b+1},$ so that $\F_s$ is a subfield of $\F_q.$ By Theorem 3.6,
 there  is an infinite family $C_m$ of additive cyclic codes of length $m\rightarrow \infty$ over $\F_{s^\ell}$ with rate $1/\ell$ and  relative distance
$\delta \ge \frac{1}{\ell}H_s^{-1}(1-1/\ell)>0.$ Let $D_m=C_m\otimes \F_{q^\ell},$ be the code obtained by extension of scalars from $\F_{s^\ell}$ to $\F_{q^\ell}.$
The dimension of $D_m$, as a code over $\F_{q^\ell}$ is at least the dimension of $C_m.$
Hence, the rate of $D_m$ is at least $ \ge 1/\ell.$ Similarly, the relative distance of the $D_m$ family is $>0.$
\end{proof}
\section{Conclusion}
In this paper, we have shown that additive $\lambda$-constacyclic codes over an extension field are asymptotically good.

This result is relevant to quantum coding theory.
 The codes constructed in  Theorem 3.4 are not always constacyclic, as observed after the proof.
 It is an open problem, of interest in its own right, to determine which QT codes give constacyclic codes by the map $\phi_B.$
A partial solution to that question is in \cite{GOS}.

 We have used random quasi-twisted codes to produce additive constacyclic codes by the map $\phi_B.$
 It is conceivable that there exist QT codes better than Varshamov-Gilbert bound, which, in turn, after taking their image by these maps,
 could provide additive constacyclic codes, or cyclic codes with a better lower bound on their relative distance.
\section{Acknowledgement}
This research is supported by National Natural Science Foundation of China (61672036), Excellent Youth Foundation of Natural Science Foundation of Anhui Province (1808085J20),
Technology Foundation for Selected Overseas Chinese Scholar, Ministry of Personnel of China (05015133) and
Key projects of support program for outstanding young talents in Colleges and Universities (gxyqZD2016008).


\begin{thebibliography}{1}

\bibitem{AGSS} A. Alahmadi, C. G\"uneri, H. Shoaib, and P. Sol\'e, ``Long quasi-polycyclic
$t$-CIS codes," {\it Advances in Math of Communication}, vol. 2, no. 1, pp. 189--198, 2018.

\bibitem{AOS} A. Alahmadi, F. \"Ozdemir, and P. Sol\'e, ``On self-dual double circulant codes," {\it Designs, Codes Cryptogr.}, DOI 10.1007/s10623-017-0393-x, 2017.

\bibitem{AM} E. F. Assmus, H. F. Mattson, and R. Turyn, ``Cyclic Codes," AF Cam-
bridge Research Labs , Bedford, MA, Summary Sci. Rep. AFCRL-66--348, 1966.

\bibitem{BM} L.M.J. Bazzi, and S.K. Mitter, ``Some randomized code constructions from group actions," {\it IEEE Trans. Inform. Theory}, vol. 52, no. 7, pp. 3210--3219, Jul. 2006.

\bibitem{B} E. R. Berlekamp,{ \em Algebraic Coding Theory}, McGraw-Hill (1968).

\bibitem{B1} J. Bierbrauer, ``The theory of cyclic codes and a generalization to additive codes," {\it Designs, Codes Cryptogr.}, vol. 25, no. 2, pp. 189--206, Feb. 2002.

 \bibitem{B2} J. Bierbrauer, {\it Introduction to Coding Theory}, Chapman and Hall/CRC Press, Boca Raton, 2005.

 \bibitem{BE} J. Bierbrauer, and Y. Edel, ``Quantum twisted codes," {\it J. Comb. Des}, vol. 8, no. 3, pp. 174--188, Apr. 2000.

 \bibitem{C+}A.R. Calderbank, E.M. Rains, N.J.A. Sloane, Quantum error correction via codes over $GF(4)$, IEEE Trans. on Inform. Theory vol. 44, pp. 1369--1387, 1998.

 \bibitem{CCC} Y. Cao, X. Chang, and Y. Cao, ``Constacyclic $\F_q$-linear codes over $\F_{q^\ell}$,"  {\it Applicable Algebra in Engineering Communication and Computing}, vol. 26, no. 4, pp. 369--388, 2015.

\bibitem{Ca} G. Castagnoli. On the asymptotic asymptotically badness of cyclic codes with block-lengths composed from a fixed set
of prime factors. In Applied algebra, algebraic algorithms and error-correcting codes, vol. 357
of Lecture Notes in Comput. Sci., pages 164--168. Springer, Berlin, 1989.

\bibitem{C} P. Charpin, Open problems on cyclic codes, p. 963--1064, in {\it Handbook of Coding Theory}, W.C. Hufman, V. Pless, eds, North-Holland, 1998.

\bibitem{CPW} C. L. Chen, W. W. Peterson, and E. J. Weldon, ``Some results on quasi-cyclic codes", {\it Information
and Control}, vol. 15, no. 5, pp. 407--423, 1969.

\bibitem{EL}S. Evra, E. Kowalsky , A. Lubotzky, ``Good cyclic codes and the uncertainty principle", {\tt https://arxiv.org/pdf/1703.01080.pdf}

\bibitem{GOO} C. G\"uneri, F. \"Ozdemir, and F. \"Ozbudak, ``Hasse-Weil bound for additive cyclic codes," {\it Designs, Codes Cryptogr.}, vol. 82, no. 1--2, pp. 249-263, 2017.

\bibitem{GOS}C. G\"uneri, F. \"Ozdemir, and P. Sol\'e, ``On the additive structure of quasi-cyclic codes'', to appear in Discret Math.

\bibitem{H} C. Hooley, ``On Artin's conjecture," {\it J. Reine Angew. Math.}, vol. 225, pp. 209--220, 1967.

\bibitem{HP} W. C. Huffman, and V. Pless, {\it Fundamentals of error correcting codes}, Cambridge University Press, 2003.

\bibitem{I} H. Imai, ``A theory of two-dimensional cyclic codes," {\it Information and Control}, vol. 34, no. 1, pp. 1--21, 1977.

\bibitem{JY}  Y. Jia, ``On quasi-twisted codes over finite fields," {\it Finite Fields and Their Applications}, vol. 18, no. 2, pp. 237--257, 2012.

\bibitem{K} T. Kasami,  ``An upper bound on $k/n$ for affine-invariant codes
with fixed $d/n,$" {\it IEEE Trans. Inform. Theory}, vol. 15, pp. 174--176, 1969.

\bibitem{RHN} R. Lidl, and H. Niederreiter, {\it Finite Fields}, Addison-Wesley, Reading, 1983.

\bibitem{LP} S. Lin, and E. Peterson, ``Long BCH codes are asymptotically bad," {\it Information and Control}, vol. 11, no. 4, pp. 445--451, 1967.

\bibitem{LS} S. Ling, and P. Sol$\acute{e}$, ``On the Algebraic Structure of Quasi-Cyclic Codes I: Finite Fields," {\it IEEE Trans. Inform. Theory}, vol. 47, no. 7, pp. 2751--2760, 2001.

    \bibitem{MW} C. Martinez-Perez, and W. Willems, ``Is the class of cyclic codes asymptotically good?,"
{\it IEEE Trans. on Inform. Theory}, vol. 52, no. 2, pp. 696--700, 2006.

\bibitem{PD} J. P. Pedersen and C. Dahl, Classification of pseudo-cyclic MDS codes,{\it IEEE Trans. on Inform. Theory}, vol. 37, pp. 365--370, 1991.

\bibitem{STGSS} M. Shi, J. Tang, M. Ge, L. Sok, and P. Sol$\acute{e}$, ``A special class of quasi-cyclic codes," {\it Bulletin  of the Austr. Math Soc.}, DOI: 10.1017/S0004972717000636, Aug. 2017.

\bibitem{SYP} M. Shi, and Y. Zhang,  ``Quasi-twisted codes with constacyclic constituent codes," {\it Finite Fields and Their Applications}, vol. 39, pp. 159--178, 2016.

\bibitem{SZS}  M. Shi, H. Zhu, and P. Sol$\acute{e}$, ``On the self-dual four-circulant codes," {\it Int. J. on Foundations of Computer Science}, to appear. {\tt https://arxiv.org/pdf/1709.07548.pdf}.

\bibitem{W} J. Wolfmann, ``A new construction of the binary golay code $(24, 12, 8)$ using a group algebra over a finite field,'' {\it Discrete Mathematics}, vol. 31, no. 3, pp. 337-338, 1980.

\end{thebibliography}
\end{document}